\documentclass[useAMS,usenatbib]{mn2e}
\usepackage{amssymb,amsmath,latexsym,graphicx,natbib,mathrsfs}

\usepackage[usenames,dvipsnames]{color}

\title[The implications of a CEW process on MSP production] 
{The implications of a companion enhanced wind on millisecond pulsar production}

\author[S. L. Smedley, C. A. Tout, L. Ferrario and D. T. Wickramasinghe]
{Sarah L. Smedley$^1$, Christopher A. Tout$^{1,2,3}$, Lilia Ferrario$^3$\\
\newauthor
and Dayal T. Wickramasinghe$^3$\\
$^1$ Institute of Astronomy, The Observatories, Madingley Road, Cambridge, CB3 0HA\\
$^2$ Monash Centre for Astrophysics, School of Mathematics, Building 28,
Monash University, Clayton, VIC 3800, Australia\\
$^3$ Mathematical Sciences Institute, The Australian National University,
ACT 0200, Australia\\
}

\begin{document}

\date{Accepted.  Received ; in original form}
\pagerange{\pageref{firstpage}--\pageref{lastpage}} \pubyear{}

\maketitle

\label{firstpage}

\begin{abstract}
The most frequently seen binary companions to millisecond pulsars (MSPs) are helium white dwarfs (He WDs). The standard rejuvenation mechanism, in which a low- to intermediate-mass companion to a neutron star fills its Roche lobe between central hydrogen exhaustion and core helium ignition, is the most plausible formation mechanism. We have investigated whether the observed population can realistically be formed via this mechanism. We used the Cambridge \textsc{stars} code to make models of Case~B RLOF with Reimers' mass loss from the donor. We find that the range of initial orbital periods required to produce the currently observed range of orbital periods of MSPs is extremely narrow. To reduce this fine tuning, we introduce a companion enhanced wind (CEW) that strips the donor of its envelope more quickly so that systems can detach at shorter periods. Our models indicate that the fine tuning can be significantly reduced if a CEW is active. Because significant mass is lost owing to a CEW we expect some binary pulsars to accrete less than the $0.1\,\rm M_{\odot}$ needed to spin them up to millisecond periods. This can account for mildly recycled pulsars present along the entire $M_{\rm c}$--$P_{\rm orb}$ relation. Systems with $P_{\rm spin} > 30\,\rm ms$ are consistent with this but too few of these mildly recycled pulsars have yet been observed to make a significant comparison.  
\end{abstract}


\section{Introduction}

At least 48\,per cent of binary millisecond pulsars host a helium white dwarf companion. Such systems are likely to have formed via the standard rejuvenation mechanism in which  a low- to intermediate-mass companion to a neutron star (NS) fills its Roche lobe between central hydrogen exhaustion and core helium ignition \citep{1982Alpar,1982Radhak}. The NS is spun up to millisecond periods by accretion from the inner edge of a disc during Roche-lobe overflow (RLOF). If the NS has a strong magnetic field, this is assumed to be buried during accretion and not to re-emerge. The NS could have formed in an earlier supernova or by accretion-induced collapse (AIC) of a white dwarf \citep{2010Hurley}. Over the last couple of decades, many models of the formation of these systems have been computed. The remnant of the donor is a white dwarf that formed as the core of the giant companion. The orbital period of the BMSP is therefore determined when the giant detaches from its Roche lobe and this depends only on the core mass. Final orbital periods of the systems range from less than $1\,\rm d$ up to around $1000\,\rm d$. In this paper we ask, ``Can the observed population of MSPs with He WD companions form from a plausible range of initial orbital periods?'' \par
\begin{figure}
\centering
\includegraphics[width=8.3cm]{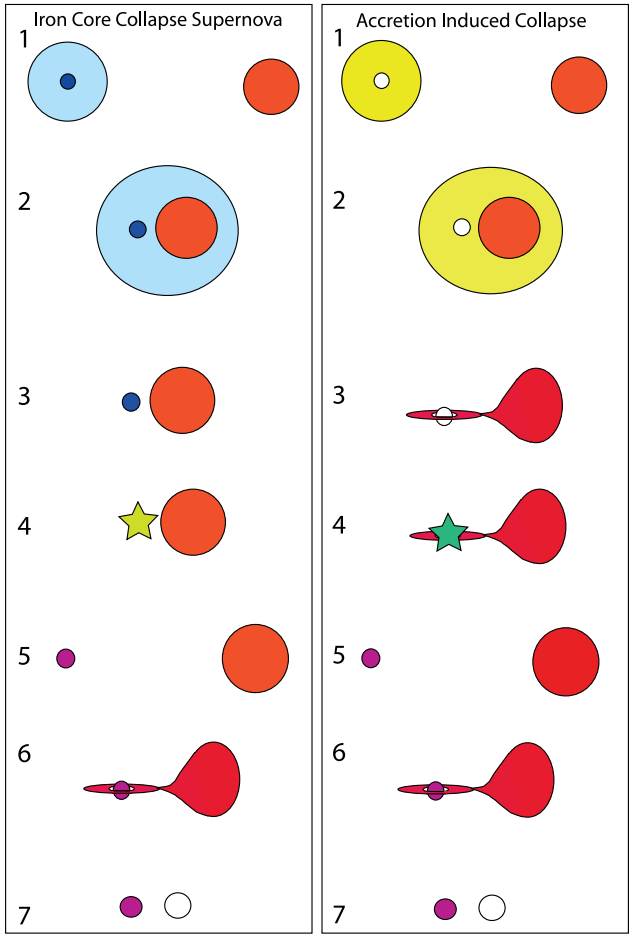}
\caption{Schematic of two formation pathways that have a MSP and a He WD pair in their final stage (not to scale). Iron core-collapse route: 1) A high-mass star in a binary system with a low-mass star. 2) The high-mass star fills its Roche lobe and a CE phase begins. 3) During the CE phase, the binary orbit shrinks and the envelope of the high-mass star is ejected leaving the massive but small helium core of the primary in a tighter orbit with its main-sequence companion. 4) The naked helium core goes on to burn iron in its centre when it undergoes a core-collapse supernova. 5) The orbit widens and leaves a NS in a close binary with its still MS companion. 6) The MS companion evolves, while magnetic braking and gravitational radiation shrink the orbit, and RLOF begins. 7) After the cessation of RLOF, the NS is left in a binary system with the core of its companion. Accretion-induced collapse route: 1) A star of $8$ to $11\,\rm M_{\odot}$ in a binary with a low-mass star, wide enough to grow an ONeMg core. 2) The higher-mass star fills its Roche lobe and a CE phase begins. 3) The orbit shrinks and the envelope of the higher-mass star is ejected leaving an ONeMg core in a tighter orbit with its MS companion. The MS star evolves and transfers matter and angular momentum to the ONeMg WD. The ONeMg WD collapses to a NS. The binary detaches from RLOF. Stages 5 to 7 for the AIC case are the same as the CC SN case and we model these stages in this work.}
\label{mech} 
\end{figure}

 The formation of BMSPs typically requires a common-envelope (CE) phase of evolution because the star that is to become the pulsar must initially have space to evolve but end up close enough to accrete from a red giant companion. \cite{2010Hurley} demonstrated that the pulsar could result either from a supernova or an AIC after the CE phase. In the CE unstable mass transfer leads to a common envelope around the two dense cores that then spiral together as the envelope is ejected \citep{2013Ivanova}. The physics involved and thence the final orbital period distribution remains uncertain. Fig.~\ref{mech} shows a schematic of both the supernova and AIC formation pathways that produce MSPs with He WD companions. In this paper, we study stages 5 to 7 of these formation pathways. These latter stages are identical for both pathways. It is the period at the start of stage 5 that we refer to as $P_{\rm orb,i}$ in what follows. However for the AIC cases we could equally assign the $P_{\rm orb,i}$ to the orbital period at the end of the CE phase.

\section{Fine tuning of the $M_{\rm c}$--$P_{\rm orb}$ Relation}
\label{fine}

We investigate the range of initial periods that produces the observed range of orbital periods of the BMSPs with He WD companions. Here initial is when the NS has just formed (via AIC or supernova) and its companion is on the main sequence (stage 5 in both pathways on Fig.~\ref{mech}). The periods at this point result from convoluted evolution involving poorly understood common-envelope evolution and supernovae kicks. We do not therefore profess to know the distribution $P_{\rm orb,i}$ but there is no reason for it not to be reasonably uniform and to cover a wide range. However on close inspection, the canonical $M_{\rm c}$--$P_{\rm orb}$ relation presented by \cite{2014Smedley} has a limitation: it gives a very large range of post-RLOF (final) orbital periods for a very small range of initial orbital periods. This is of concern because an incredibly fine-tuned initial configuration is less likely to be a realistic model. 
\par
\begin{figure}
\centering
\includegraphics[width=8.9cm]{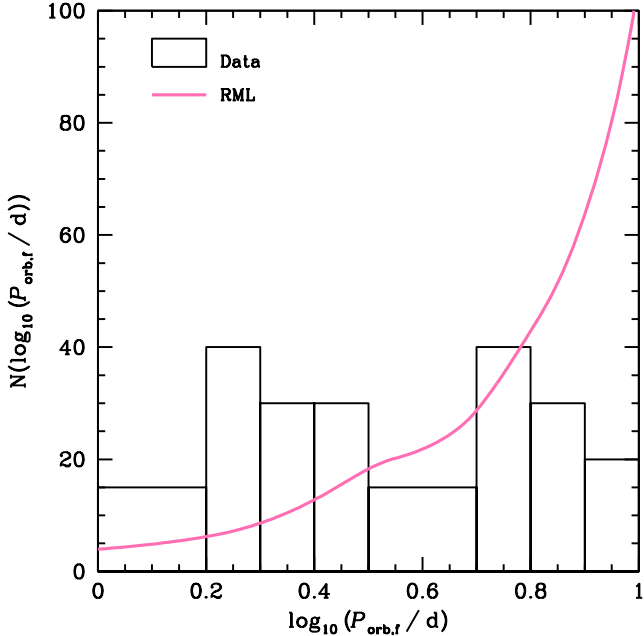}
\caption[Comparison of the final period distributions for data and $B=0$]{The logarithmic number density plotted against $\log_{10}\,(P_{\rm orb,f})$ of the systems in the ATNF Pulsar Catalogue with He WD companions for $1 < P_{\rm orb,f}\,/\,\rm d < 10$. The pink solid line is the expected final period distribution when there is RML from the donor. Notice that the actual distribution appears much flatter than the prediction.}
\label{pdist1} 
\end{figure}

 Fig.~\ref{pdist1} shows the distribution of the orbital periods of the observed MSPs with He WD companions from the ATNF Pulsar catalogue. These systems are listed in the Appendix of \cite{2014Smedley}. The selection criteria applied to that data set were also used in this calculation. After the selection cuts 25 systems remain, with orbital periods between $1$ and $10\,\rm d$, plotted on a logarithmic scale. The period distribution of the data is rather flat in $\log_{10}(P_{\rm orb,f})$ except for fewer systems below $\log_{10}(P_{\rm orb,f}\,/\,\rm d) = 0.2$ and a dip of comparable size around $\log_{10}(P_{\rm orb,f}\,/\,\rm d) = 0.6$. Also plotted in Fig.~\ref{pdist1} is the final period distribution that arises when a population of binary systems, with initial donor star masses of $1\,\rm M_{\odot}$, initial NS masses of $1.55\,\rm M_{\odot}$ and initial periods distributed uniformly in $\log_{10}(P_{\rm orb,i})$, are formed via the standard rejuvenation mechanism with the donor stars losing mass according to the RML rate \citep{1975Reimers}. The final period distribution from $1$ to $10\,\rm d$, when the NS retains all the mass transferred to it, with RML from the donor, is covered by systems with initial periods from $0.94$ to $1.23\,\rm d$, a range of less than $0.3\,\rm d$. Specifically we calculated the final periods for a sample of $10^{6}$ systems with initial periods distributed uniformly in $\log_{10}(P_{\rm orb,i})$. We divided these final periods into $100$ bins in $\log_{10}(P_{\rm orb,i})$ and calculated the number density of for each bin and normalised these to the data. The pink line in Fig.~\ref{pdist1} maps out the positions of the midpoints of the tops of the $100$ bins. It shows the predicted final period distribution if these systems were formed via the standard rejuvenation mechanism with RML from the donor. There are too few systems with final periods with $0 \leq \log_{10}(P_{\rm orb,f}\,/\,\rm d) \leq 0.5$ and too many with $0.5 \leq \log_{10}(P_{\rm orb,f}\,/\,\rm d) \leq 1$. 

\par Such fine tuning occurs because a giant transferring mass to a NS continues to evolve as it does so. Even if the giant has a rather small core mass when RLOF begins, its core can grow substantially by the time it detaches. In standard models mass transfer is driven by the evolution of the giant and so such core growth is inevitable. To alleviate this fine tuning we seek a process that can drive mass evolution on a shorter timescale than that at which the giant evolves.

\section{A companion enhanced wind (CEW)}

The enhancement of tidal winds from cool subgiants/giants in interacting binary systems can play a significant role in orbital evolution. CEW was first proposed by \cite{1987Tout} to explain why the binary star Z~Her (among other RS CVn systems) comprises a more evolved star as the less massive component of the system. Z Her is composed of two subgiants (spectral classes F and K) in a $3.99\,\rm d$ orbit. They found that RML rate is a factor of 50 smaller than required to produce the Z Her system before RLOF. Reimers' mass-loss formula does predict white dwarf masses consistent with observations but it is usually believed that mass loss is stronger during AGB evolution and weaker during the RGB phase \citep{1983Iben}. \cite{1987Tout} proposed that the presence of a companion enhances the mass-loss rate in very close systems to produce the observed mass discrepancy. The nature of the enhancement is rooted in tidal friction and the dynamo activity of the star but it is not physically understood. This tidal friction becomes progressively important as the giant fills its Roche lobe. The point at which there is no RLOF before the formation of a WD is critical for the production of MSPs via rejuvenation because if there is no RLOF then there is no NS spin up. We incorporated a companion enhanced wind in our models for the production of MSPs with He WD companions to see if the enhanced mass loss from the donor can relax the fine tuning of the relation. 

\section{The Code} 
\label{trapcode}
We use a version of the Cambridge STARS code \citep{1971Eggleton,1995Pols} updated by \citet{2009Stan}.  The code features a non-Lagrangian mesh. Convection is according to the mixing-length theory of \citet{1958Bohm} with $\alpha_{\rm MLT}=2$ and convective overshooting is included as described by \citet{1997Sch}.  The nuclear species $^{1}\rm H$, $^{3}\rm He$, $^{4}\rm He$, $^{12}\rm C$, $^{14}\rm N$, $^{16}\rm O$ and $^{20}\rm Ne$ are evolved in detail. Opacities are from the OPAL collaboration \citep{1996Iglesias} supplemented with molecular opacities of \citet{1994Alex} and \citet{2005Ferg} at the lowest temperatures and by \citet{1976Buchler} at higher temperatures.  Electron conduction is that of \citet{1969Hub} and \citet{1970Can}.  Nuclear reaction rates are those of \citet{1988Cau} and the NACRE collaboration \citep{1999Angulo}. We include the rate of change of the angular momentum $\dot{J}$ by gravitational radiation \citep{1959Landau},
\begin{equation}\label{eq:gr}
\left( \frac{\dot{J}}{J} \right)_{\rm gr} = - \frac{32 G^{3}}{5 c^{5}} \left( \frac{M_{\rm a} M_{\rm d} M_{\rm B}}{a^{4}} \right),
\end{equation}
where $M_{\rm a}$ is the mass of the accretor, $M_{\rm d}$ the mass of the donor, $M_{\rm B} = M_{\rm d} + M_{\rm a}$ the mass of the binary system and $a$ its semi-major axis. We include magnetic braking at the empirical rate of \citet{1981Verbunt},
\begin{equation}\label{eq:mb}
\left( \frac{\dot{J}}{J} \right)_{\rm mb} = -0.5 \times 10^{-28}\,\rm s^{2}\,\rm cm^{-2} f_{\rm mb}^{-2} \frac{IR_{\rm d}^{2}}{a^{5}} \frac{GM_{\rm B}^{2}}{M_{\rm a}M_{\rm d}}\quad \rm s^{-1},
\end{equation}
where $R_{\rm d}$ is the radius of the donor star and $I$ its moment of inertia. The constant factor $f_{\rm mb}$ was chosen to fit the equatorial velocities of G~and K~type stars \citep{smith1979}.
\par A CEW was implemented by a multiplicative factor applied to single star mass loss which can be conveniently described with Reimers' formula, 
\begin{equation}\label{eq:reem}
\dot{M}_{\rm R} = -4 \times 10^{-13}\,\eta\,\frac{L}{\,\rm L_{\odot}}\frac{R}{\,\rm R_{\odot}}\frac{\,\rm M_{\odot}}{M}\,\rm M_{\odot}\,\rm yr^{-1},
\end{equation}
where $\eta$ is a free parameter taken to be 1 for this work. However $\eta$ is believed to vary depending on the stage of evolution of the star. We used the CEW enhancement proposed by \cite{1987Tout}. It takes the form 
\begin{equation}\label{eq:trap}
\dot{M}_{\rm CEW} = \dot{M}_{\rm R} \left( 1 + B \min \left[\left( \frac{R_{\rm d}}{R_{\rm L}}\right)^{\!\!\!6}, \frac{1}{2^{6}} \right] \right),
\end{equation}
 where $B$ is a strength parameter, $R_{\rm d}$ is the radius of the donor star and $R_{\rm L}$ is its Roche lobe radius. When $B=0$ this reduces to the original RML. We note that this relation was not derived from physical or observed empirical laws. It is a prescription proposed by \cite{1987Tout} and used to fit observations of systems like Z Her. The saturation at $R = \frac{1}{2} R_{\rm L}$ is included because the star is expected to be in full corotation by then and differential rotation that can drive a dynamo is controlled. The angular momentum loss in the CEW is
\begin{equation}\label{eq:crew}
\left( \frac{\dot{J}}{J} \right)_{\rm CEW} = \frac{\dot{M}_{\rm CEW} M_{\rm d}}{M_{\rm a}M_{\rm B}},
\end{equation}
which assumes the wind carries off all the specific angular momentum of the donor. In addition to this non-conservative enhanced wind we also allow for inefficient RLOF in our models to account for any mass lost from compact objects such as by the NS in a propeller mechanism \citep{1975Illarionov} during RLOF. The propeller mechanism is only operational at high-spin rates once the pulsar has spun up. This was included in the \textsc{stars} code by defining the efficiency parameters $\alpha$, the fraction of mass transferred by the donor that reaches the accretor and $\beta$, the fraction of this actually captured by the accretor over long timescales.  Lost material carries away the specific angular momentum of the component from which it is actually lost so that
\begin{equation}\label{eq:eff}
\left( \frac{\dot{J}}{J} \right)_{\rm RL} = \frac{\dot{M_{\rm d}}}{M_{\rm d}} \left[ \frac{1- \alpha + \alpha (1-\beta) q^{2}}{1+q} \right],
\end{equation}
where $M_{\rm d}/M_{\rm a} = q$. Here we keep $\alpha = 1$ so that transferred material is only lost after reaching the accretor if $\beta < 1$. We varied $\beta$ to explore the effects of varying the amount of transferred matter that is retained by the accretor and hence not lost in mass ejections. The fraction $\beta$ may well vary over time but here we test three different efficiencies and keep them constant throughout their respective evolutions to simplistically simulate mass lost from the NS in mass ejections of different magnitudes. The total angular momentum evolution, 
\begin{equation}
\left( \frac{\dot{J}}{J} \right)_{\rm tot} = \left( \frac{\dot{J}}{J} \right)_{\rm mb} + \left( \frac{\dot{J}}{J} \right)_{\rm gr} +\left( \frac{\dot{J}}{J} \right)_{\rm RL} + \left( \frac{\dot{J}}{J} \right)_{\rm CEW},
\end{equation}
includes the effects of magnetic braking, gravitational radiation, Roche lobe overflow and the companion enhanced wind.

\section{Detailed Models including a CEW}

To investigate the CEW, we compute models of the evolution of a $1\,\rm M_{\odot}$ donor star with a $1.55\,\rm M_{\odot}$ NS companion through a phase of RLOF. The evolution of the donor star is fully computed and the NS is treated as an accreting point mass. The donor star begins its evolution as a zero-age main-sequence star in thermal equilibrium and we assume that the NS was formed prior to the starting point either by an AIC of an ONeMg WD \citep{2007Lilia,2010Hurley} or a core-collapse supernova of a high-mass star. We follow the evolution up to and through an X-ray binary phase until the companions detach as a MSP with a He WD companion. This is stages 5 to 7 in Fig.~\ref{mech}.

\subsection{Fine tuning at Low Periods}
\label{rlofeff} 

As emphasised in Section~\ref{fine}, for the case where there is only RML from the donor, the $P_{\rm orb,i}$--$P_{\rm orb,f}$ relation is finely tuned at short periods. The fine tuning occurs because the mass transfer is driven by the evolution of the giant and so proceeds on a similar timescale as core growth. Thus, even when the RLOF begins when the core is small, the core can grow substantially by the end of RLOF. To investigate the effect that a CEW has on the relationship between the initial and final period of the system, we made models along the entire expanse of Case~B RLOF for $B=0, 100, 300, 500$ and $1000$. Only systems that underwent Case~B RLOF were used so as to exclude donors that may not leave helium cores when the system detaches after RLOF. Higher $B$s were not considered at this time but we discuss them further in Section~\ref{nonmspssec}. Three different levels of RLOF efficiency, $\beta = 0.5, 0.75$ and $1$, were tested to include possible mass loss from the NS. 

\begin{figure}
\centering
\includegraphics[width=8.9cm]{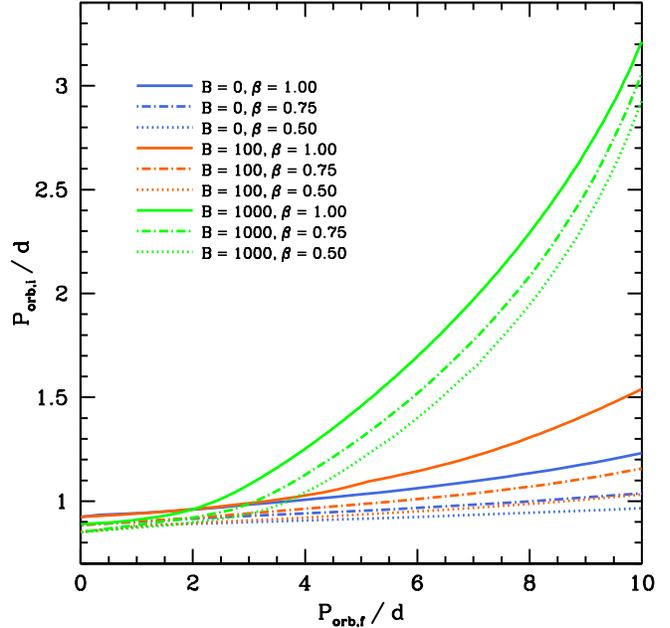}
\caption{Plot to show the relationship between the orbital periods at the formation of the NS $P_{\rm orb,i}$ and those of the millisecond pulsars with WD companions $P_{\rm orb,f}$ after the end of RLOF. There are three different strengths of CEW plotted here for three different RLOF efficiencies ($\beta = 1, 0.75$ and $0.5$). The blue lines represent the $B=0$ cases, the orange lines the $B=100$ and the green lines $B=1000$.}
\label{finetuneclose} 
\end{figure}

Fig.~\ref{finetuneclose} shows the tracks for the three different $B$s through the $P_{\rm orb,i}$--$P_{\rm orb,f}$ plane for three different $\beta$s at short periods ($P_{\rm orb,f} \leq 10\,\rm d$). The final periods are shorter for a given initial period than those found by \cite{2014Smedley}. This is because, in this study, we have included angular momentum loss owing to the emission of gravitational radiation and mass loss from the donor in winds. 
Considering the case when all of the mass transferred to the NS is retained ($\beta = 1$, solid lines), the models show that fine tuning at shorter periods can be significantly reduced if a CEW is active in the systems. The CEW removes the mass at a faster rate, revealing the WD core of the donor before it has grown and so at a shorter orbital period. When $B$ is large the mass evolution is driven by the wind \citep{1991Tout} on a timescale shorter than the nuclear timescale on which the core grows. Specifically for our donors the nuclear timescale for expansion is about 10 times the mass loss timescale when $R \approx \frac{1}{2} R_{\rm L}$ and $B=1000$. At $B=1000$ the $P_{\rm orb,i}$--$P_{\rm orb,f}$ relations are much steeper and give a narrower range of final periods for a given range of initial periods. Comparison of the the different $\beta$ cases shows that a decrease in the efficiency of the mass transfer keeps the $P_{\rm orb,i}$--$P_{\rm orb,f}$ relation finely tuned at the low core masses even as $B$ increases. To demonstrate the implications of fine tuning, ponder the case where $\beta=0.5$. Here the $P_{\rm orb,i}$--$P_{\rm orb,f}$ relation for $B=0$ has a very shallow gradient in initial period up to a final orbital period of $25\,\rm d$. According to the ATNF pulsar catalogue, about 63 per cent of the MSPs with He WD companions observed in the Milky Way have periods between $1$ and $25\,\rm d$. Thence if $B=0$ and $\beta = 0.5$ in reality, just under two thirds of the population of MSPs with He WD companions would originate from a minuscule range in initial period of around $0.2\,\rm d$.

\begin{figure}
\centering
\includegraphics[width=8.8cm]{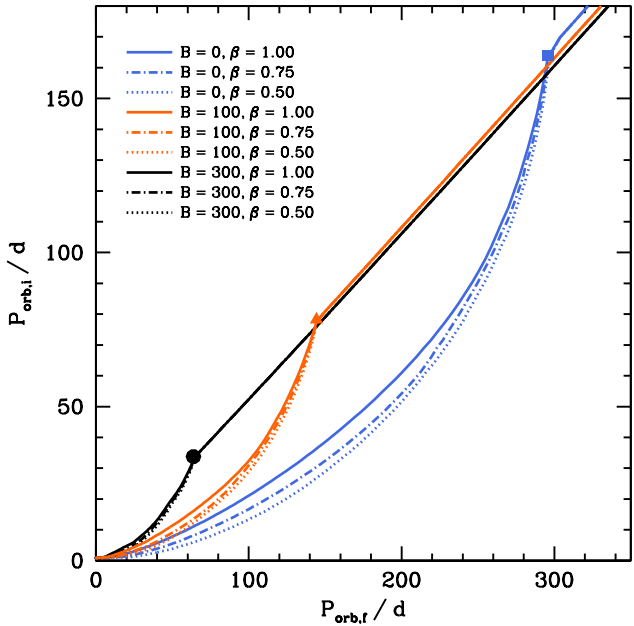}
\caption{Plot to show the relationship between the orbital period at the beginning and end of evolution for the cases where $\beta=1, 0.75$ and $0.5$. There are nine cases plotted here. The blue lines all have $B=0$, the orange lines $B=100$ and black $B=300$. The solid lines all have $\beta=1.00$, the dot-dashed lines $\beta=0.75$ and dotted $\beta=0.50$. The circle represents the point $P_{\rm i, crit}$ in $P_{\rm orb,i}$--$P_{\rm orb,f}$ space where the $B=300$ cases no longer have a period of RLOF before the donor becomes a WD, the triangle is $P_{\rm i,crit}$ for the $B=100$ cases and the square is $P_{\rm i,crit}$ for $B=0$ cases.}
\label{finetunewide1} 
\end{figure}

\subsection{Exploring $P_{\rm orb,i}$--$P_{\rm orb,f}$ Space}

Now for a larger range of $P_{\rm orb,i}$--$P_{\rm orb,f}$ space, Fig.~\ref{finetunewide1} shows three cases of $B$ for three cases of $\beta$, nine cases in total. For each $B$ there is an initial period above which there is no RLOF because the envelope is lost before the giant has grown sufficiently large. This initial period $P_{\rm i,crit}$ is smaller for larger $B$ \citep{1987Tout}. Thus $P_{\rm i, crit}$ puts a limit on the initial periods that produce MSP binary systems with He WDs for a given $B$ and also on the final periods for which various $B$s can produce MSPs. For $P_{\rm orb,i} > P_{\rm i,crit}$ there is no RLOF so there is no transfer of mass and angular momentum to the NS via RLOF so, crucially, there are no rejuvenated MSPs produced in such systems because spin up of the MSP is an essential component of the formation pathway. For $B=500$ with $\beta=1$, $P_{\rm i,crit} = 17\,\rm d$ and for $B=100$ with $\beta=1$, $P_{\rm i,crit} = 78\,\rm d$. These correspond to final periods of $33\,\rm d$ and $149\,\rm d$ respectively. The coverage of the $P_{\rm orb,i}$--$P_{\rm orb,f}$ plane is different for each $B$.

\begin{figure}
\centering
\includegraphics[width=8.8cm]{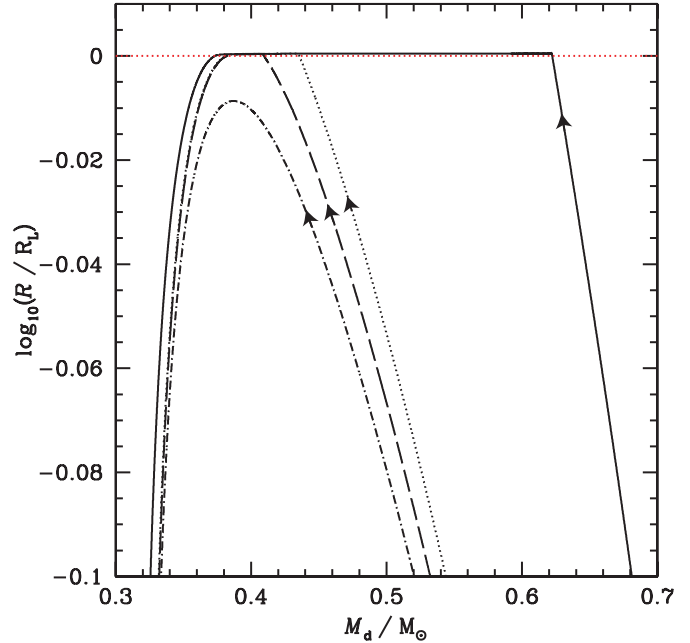}
\caption{Behaviour when $P_{\rm orb,i} \approx P_{\rm i,crit}$. Four cases are shown with different initial periods but all with $B=100$ and $\beta=1$. When $\log_{10}\,(R\,/\,R_{\rm L}) \geq 0$ RLOF takes place and when $\log_{10}\,(R\,/\,R_{\rm L}) < 0$ it does not. Here $R = R_{\rm L}$ is marked with a red dotted line. The solid line has an initial period of $50\,\rm d$, the black dotted line $75\,\rm d$, the dashed line $77\,\rm d$ and the dot-dashed line $79\,\rm d$. When $P_{\rm orb,i} = 79\,\rm d$ there is no RLOF at all. For $P_{\rm orb,i} = 75$ and $77\,\rm d$ there is a small amount of matter accreted by the NS. At $P_{\rm orb,i} = 50\,\rm d$ there is sufficient mass transfer to spin up the NS to millisecond periods. All of these systems have initial donor star masses of $1\,\rm M_{\odot}$.}
\label{rlofvsm} 
\end{figure}

Fig.~\ref{rlofvsm} illustrates how binary systems with $P_{\rm orb,i}$ close to $P_{\rm i, crit}$ evolve, for $B=100$ and $\beta=1$ (close to the orange triangle in Fig.~\ref{finetunewide1}). The radius of the star relative to its Roche lobe radius $\log_{10}\,(R\,/\,R_{\rm L})$ is plotted against the mass of the donor star $M_{\rm d}$. The binary evolution of four systems with initial orbital periods of $50$, $75$, $77$ and $79\,\rm d$ is shown. The system with $P_{\rm orb,i} = 79\,\rm d$ does not engage in RLOF at any point during its evolution so a NS in such a system would not be spun up to millisecond periods. Thus systems with $P_{\rm orb,i} > P_{\rm i,crit}$ do not become MSPs. For initial orbital periods of $75\,\rm d$ and $77\,\rm d$ only a small fraction of the required $0.1\,\rm M_{\odot}$ needed to spin up the NS to millisecond periods \citep{spinup} is accreted. The system with $P_{\rm orb,i} = 50\,\rm d$ does engage in RLOF for long enough to allow $0.1\,\rm M_{\odot}$ of matter to be accreted by the NS. As RLOF efficiency decreases so the RLOF episode must increase to allow the NS to capture sufficient mass. Hence as $P_{\rm i,crit}$ decreases fewer systems become MSPs. We show in Section~\ref{rlofeff} that the decline of RLOF efficiency increases the fine tuning at lower periods. So the lower the RLOF efficiency the larger the range of final periods for a given range of initial periods. For $B=100$ with $\beta=1$, $P_{\rm i, crit} = 78\,\rm d$ and the maximum initial period that is in RLOF long enough for the NS to accrete $0.1\,\rm M_{\odot}$ is $P_{0.1\,\rm M_{\odot}} = 58\,\rm d$. 

\begin{figure}
\centering
\includegraphics[width=8.9cm]{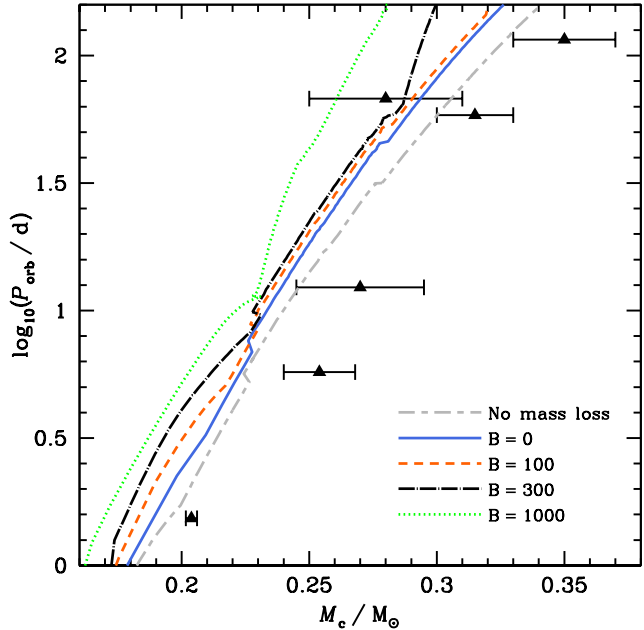}
\caption[The effect of B on the $M_{\rm c}$--$P_{\rm orb}$ relation.]{Plot to show how the $M_{\rm c}$--$P_{\rm orb}$ relation depends on $B$ when a CEW is left active after the cessation of RLOF. The solid triangles are the systems with $P_{\rm orb} > 1\,\rm d$ and their respective error bars. There are 5 $M_{\rm c}$--$P_{\rm orb}$ relations plotted.}
\label{porbmcrel} 
\end{figure}

\subsection{The $M_{\rm c}$--$P_{\rm orb}$ Relation}

For donor stars of the masses considered here, the luminosity depends only on the helium core mass and the radius depends only on the luminosity and the total mass \citep{pacz} so there is a relation between the orbital period $P_{\rm orb}$ and the remnant mass $M_{\rm c}$ when the system detaches \citep{1971Refsdal}. The $M_{\rm c}$--$P_{\rm orb}$ relation does not depend on how a system gets to this point. Neither the efficiency of mass transfer, the initial NS mass nor the initial donor star mass affect the relation. The metallicity of the donor star does have an affect but relations remain consistent with the data at all reasonable metallicities. We might not expect $B$ to affect the $M_{\rm c}$--$P_{\rm orb}$ relation much but, because the CEW continues after the remnant detaches, there is a small effect. Five $M_{\rm c}$--$P_{\rm orb}$ relations are plotted in Fig.~\ref{porbmcrel}, four with mass loss in winds (each with a different $B$) and one with no mass loss in a wind at all. MSPs with He WD companions that have measured masses are plotted too. Table~\ref{pulsartable2} lists these selected BMSPs with measured pulsar and companion masses in orbital periods greater than $1\,\rm d$. Our selection criteria exclude, 1) systems in globular clusters, 2) pulsars with $P_{\rm spin} > 30\,\rm ms$, 3) minimum companion masses greater than $0.5\,\rm M_{\odot}$, 4) systems with orbital periods less than $1\,\rm d$, 5) multiple systems and 6) systems with companions known to be not He WDs. 
\begin{table*}
\centering
\caption{BMSPs with measured pulsar and companion masses in orbital periods greater than $1\,\rm d$. \label{pulsartable2}}\vspace{5pt}
\begin{tabular}{| l | r | r | l | l | r ||}
\hline
Name & $P_{\rm spin} /\rm ms$ & $P_{\rm orb}/\rm d$ & $M_{\rm p}/\rm M_{\odot}$ & $M_{\rm comp}/\rm M_{\odot}$ & Reference \\
\hline
\hline
J1909--3744 &  2.95 &    1.53        & $1.438\pm 0.024$& $0.2038 \pm 0.0022$ & \cite{2005Jacoby}     \\
J0437--4715 &  5.76 &    5.74        & $1.76 \pm 0.02$ & $0.254  \pm 0.014$  & \cite{2008Verbiest}   \\
B1855+09    &  5.36 &   12.33        & $1.6  \pm 0.2 $ & $0.270  \pm 0.025 $ & \cite{2004Splaver}    \\
J1910+1256  &  4.99 &   58.47        & $1.6  \pm 0.6$  & $0.30 - 0.33$       & \cite{2011Gonzalez}   \\
J1713+0747  &  7.99 &   67.83        & $1.3  \pm 0.2$  & $0.28   \pm 0.03$   & \cite{2005Splaver}    \\
J1853+1303  &  4.09 &  115.65        & $1.4  \pm 0.7$  & $0.33 - 0.37$       & \cite{2011Gonzalez}   \\
\hline  
\end{tabular}
\end{table*}

All of the plotted relations in Fig.~\ref{porbmcrel} exhibit a jump around $M_{\rm c} = 0.225$. This was noted by \cite{2014Smedley} and \cite{2014Jia} found that it is due to a temporary contraction of the donor star when the H-burning shell crosses the hydrogen discontinuity. There is a second jump common to all the relations at $M_{\rm c} = 0.28$. Fig.~\ref{porbmcrel} shows that when the CEW is left active for the entire evolution, including time after RLOF has ceased, the $M_{\rm c}$--$P_{\rm orb}$ relation is affected by $B$. However when mass loss is switched off at the cessation of RLOF, the relations all converge to the `No mass loss' relation. The models used to construct the `No mass loss' relation had no mass loss from the system neither from stellar winds nor non-conservative RLOF. This is the same as the canonical relation of \cite{2014Smedley}. If the CEW remains operational post-RLOF, additional stripping of the white dwarfs' envelopes takes place because the enhanced mass loss remains active over the brief phase in which the giant envelope contracts from $R = R_{\rm L}$ to $R=R_{\rm WD}$ on a thermal timescale. So the $M_{\rm c}$--$P_{\rm orb}$ relation moves upwards and to the left. Even if only a few hundredths of a solar mass are stripped off after the cessation of RLOF, the $M_{\rm c}$--$P_{\rm orb}$ relation is noticeably altered. We note that there is no reason to disable the CEW after the cessation of RLOF, so that a leftward shift in the $M_{\rm c}$--$P_{\rm orb}$ plane is expected to exist and grow with time. Since all $M_{\rm c}$--$P_{\rm orb}$ relations plotted in Fig.~\ref{porbmcrel} fit the available data reasonably well, none of them can be ruled out. They do however agree less well with the measured masses as mass loss by the CEW increases. 

\subsection{The Neutron Star Mass}

\cite{2014Smedley} showed that the MSPs in binary systems with He WD companions, when formed by the standard rejuvenation mechanism, are likely to have masses between $1.55$ and $1.65\,\rm M_{\odot}$. Masses of newly formed NSs are thought to be about $1.25\,\rm M_{\odot}$ if produced via an AIC of an ONeMg WD or $1.35\,\rm M_{\odot}$ if formed via a core-collapse supernova \citep{2010Schwab}. The $M_{\rm c}$--$P_{\rm orb}$ relation is not sensitive to the initial mass of the NS. However the initial NS mass becomes important when studying the $P_{\rm orb,i}$--$P_{\rm orb,f}$ relation. \cite{2014Smedley} showed that lower-mass NSs end up with lower-mass companions than their higher-mass counterparts. Therefore the lower-mass NS systems have larger initial orbital periods for a given core mass. \par
\begin{figure}
\centering
\includegraphics[width=8.9cm]{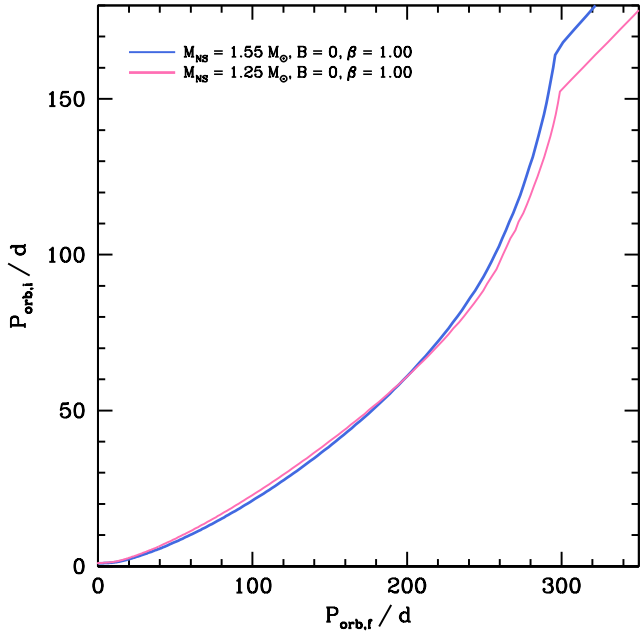}
\caption{Plot of the $P_{\rm orb,i}$--$P_{\rm orb,f}$ relations for two different NS masses. The case where the initial NS mass was $1.55\,\rm M_{\odot}$ is plotted in blue and the case where the initial NS mass was $1.25\,\rm M_{\odot}$ in pink. The difference between the $P_{\rm orb,i}$--$P_{\rm orb,f}$ relationship exhibited by the two NS masses is small.}
\label{nsmass} 
\end{figure}

 We compute the $P_{\rm orb,i}$--$P_{\rm orb,f}$ relation for initially low-mass ($1.25\,\rm M_{\odot}$) NSs and medium-mass ($1.55\,\rm M_{\odot}$) NSs to demonstrate the effect of varying NS mass. Fig.~\ref{nsmass} shows two $P_{\rm orb,i}$--$P_{\rm orb,f}$ relations when $B=0$, one where the initial NS mass is $1.55\,\rm M_{\odot}$ and the other $1.25\,\rm M_{\odot}$. Fig.~\ref{nsmass} shows that the two curves are nearly identical up to $P_{\rm orb,i} = 60\,\rm d$ but the lower-mass NS systems have slightly larger initial orbital periods for a given final period. This behaviour is reversed at higher periods. When $B=1000$ the $P_{\rm orb,i}$--$P_{\rm orb,f}$ relations for these same two initial NS masses remain very similar. Thus the initial NS masses across all $B$s studied here have an insignificant effect on the $P_{\rm orb,i}$--$P_{\rm orb,f}$ relation. 

\par We have found that we can achieve a better fit with observations if the NS masses are in the range $1.55$ to $1.65\,\rm M_{\odot}$ at the end of RLOF. This range is more likely to be reached if we use an initial NS mass of $1.25\,\rm M_{\odot}$. Considering an initial donor mass of $1\,\rm M_{\odot}$, the minimum amount of mass it must lose to become a He WD is about $0.5\,\rm M_{\odot}$\footnote{The maximum mass of a core before the He flash is $0.472\,\rm M_{\odot}$ but $0.5\,\rm M_{\odot}$ is used here for a simple demonstration.}. The maximum amount the donor can lose is about $0.85\,\rm M_{\odot}$\footnote{Limited by the Schoenberg-Chandrasekhar mass which is associated with the minimum mass of a He WD, which is about $0.15\,\rm M_{\odot}$.}. Considering initial NS masses of $1.25\,\rm M_{\odot}$ and above, the case where the NS retains all the transferred matter always produces larger NS masses than the likely range $1.55$ to $1.65\,\rm M_{\odot}$. A $1.25\,\rm M_{\odot}$ NS must accrete between $0.3$ and $0.4\,\rm M_{\odot}$ to grow up to these masses. This can be used to put rough limits on the range of $\beta$ compatible with the production of these systems. For the minimum donor mass loss case (which produces the highest mass He WDs), $0.5\,\rm M_{\odot}$ must be lost from the donor in RLOF and $0.3$ to $0.4\,\rm M_{\odot}$ can be accreted with $\beta$ in the range $0.6$ to $0.8$. Likewise, the maximum donor mass loss case (which produces the lowest mass He WDs), is $0.85\,\rm M_{\odot}$ with $0.3$ to $0.4\,\rm M_{\odot}$ accreted by the NS corresponding to $\beta$ in the range $0.35$ to $0.47$. 
\par Remember that these close binary systems suffer two kinds of mass loss. Not only do they lose mass from RLOF but also continuously from the donor owing to the CEW. \cite{1991Tout} deduced that, when the timescale for mass loss is greater than the timescale on which the nuclear evolution drives the expansion of the donor, the mass-transfer rate by RLOF and the mass-loss rate by a CEW are of the same order. Mass loss can stably drive mass transfer because the mass losing giant expands faster than its Roche lobe grows owing to widening of the orbit. In our models this slows envelope growth even after RLOF has begun.

\section{The Final Orbital Period Distribution}

\begin{figure}
\centering
\includegraphics[width=8.9cm]{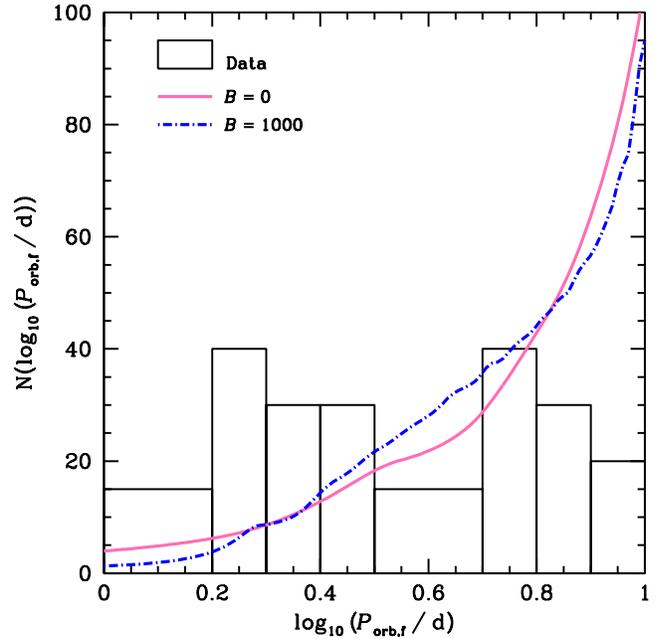}
\caption{The number density plotted against $\log_{10}\,(P_{\rm orb,f})$ of the systems in the ATNF Pulsar Catalogue with He WD companions for $1 < P_{\rm orb,f}\,/\,\rm d < 10$. The pink solid line is the expected final period distribution with RML from the donor ($B=0$). The blue dot-dashed line is the expected final distribution for $B=1000$. Note that the distributions for $B=0$ and $1000$ are very similar.}
\label{pdist2} 
\end{figure}

In Section~\ref{fine} the final orbital period distribution for a population of binary systems, formed via the standard rejuvenation mechanism with the donor stars losing mass according to RML rate was calculated. For this we used initial donor star masses of $1\,\rm M_{\odot}$, initial NS masses of $1.55\,\rm M_{\odot}$ with initial periods, at the formation of the NS, selected from a uniform distribution in $\log_{10}(P_{\rm orb,i})$ between $1$ and $10\,\rm d$. The same calculation was made with $B=1000$ and the final distribution is plotted as the blue dot-dashed line in Fig.~\ref{pdist2}. 

Overall the $B=1000$ final period distribution is similar to the final period distribution for $B=0$. However Fig.~\ref{pdist2} does not take into account the relative numbers of systems in each initial period range. The final period distribution from $1$ to $10\,\rm d$ for $B=1000$ is covered by systems with initial periods from $0.91$ to $3.21\,\rm d$, a range of about $2.3\,\rm d$. For $B=0$, initial periods from $0.94$ to $1.23\,\rm d$, a range of about $0.3\,\rm d$, cover the same final periods. The range in initial period needed to make systems with final periods in the range of $1$ to $10\,\rm d$ is about 8 times larger for $B=1000$ than for $B=0$. For a sample of systems, 23 times more fall in the initial period range needed to make systems with final periods in the range of $1$ to $10\,\rm d$ for $B=1000$ than $B=0$, assuming that the initial period is uniformly distributed in $\log_{10}(P_{\rm orb,i})$. Thus when a CEW is operational, more systems that will become MSPs form at shorter periods.

\section{Pulsars with $P_{\rm spin} > 30\,\rm ms$ and He WD Companions}
\label{nonmspssec}

Different wind enhancement $B$ leads to different $P_{\rm i,crit}$, the period above which there is no RLOF before a WD is produced. So if the CEW varies from system to system there should be mildly recycled pulsars throughout the expanse of the $M_{\rm c}$--$P_{\rm orb}$ relation. Systems with $30\,\rm ms < P_{\rm spin} < 500\,\rm ms$ are mildly recycled. Those with $10\,\rm ms < P_{\rm spin} < 30\,\rm ms$ are recycled and with $P_{\rm spin} < 10\,\rm ms$ are fully recycled. Up to now we have only considered $B < 10^{3}$ because systems with $10^{3} < B < 10^{4}$ only produce MSPs for a small range of final periods. For $B \geq 10^{4}$ the stars do not undergo RLOF at all before the donor stars become WDs so no MSP binaries are produced.

\begin{figure}
\centering
\includegraphics[width=8.9cm]{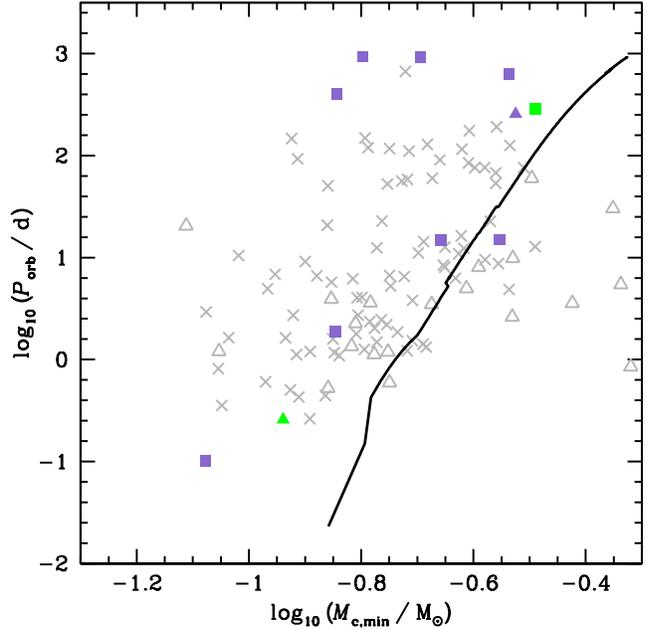}
\caption[Pulsars with $P_{\rm spin} > 30\,\rm ms$ in the $M_{\rm c}$--$P_{\rm orb}$ plane.]{Plot to show minimum masses of the pulsars with $P_{\rm spin} > 30\,\rm ms$ in binary systems with He WDs from the ATNF pulsar catalogue. The mass $M_{\rm c, min}$ is the minimum companion mass for an orbital inclination of $i=90^{\circ}$ with a MSP mass of $1.35\,\rm M_\odot$. The 9 mildly recycled pulsars are plotted in purple and the 2 normally-spinning pulsar systems are green. Squares indicate that the companion type is known to be He WD and the triangles are the two systems with unknown companion type. The grey crosses are recycled MSPs ($P_{\rm spin} < 30\,\rm ms$) in binary systems with He WDs and the grey open triangles are MSPs with unidentified companion types. The black solid line is the $M_{\rm c}$--$P_{\rm orb}$ relation of \cite{2014Smedley}.}
\label{nonmsphewds} 
\end{figure}

\begin{table*}
\centering
\caption{Binary pulsars from the ATNF pulsar catalogue with $P_{\rm spin} > 30\,\rm ms$. \label{pulsartable3}}\vspace{5pt}
\begin{tabular}{| l | r | r | l | l | l |}
\hline
Name & $P_{\rm spin} /\rm s$ & $P_{\rm orb}/\rm d$ & $M_{\rm comp, min}/\rm M_{\odot}$ & Comp. type & Reference \\
\hline
\hline
J0348+0432  & 0.0391226569017806       & 0.102424062722     & 0.083853  & He & \cite{antoniadis2013} \\
J1232-6501  & 0.0882819082341          & 1.86327241         & 0.142552  & He & \cite{2001Edwards} \\
J1711-4322  & 0.1026182883472          & 922.4708           & 0.202270  & He & \cite{2006Lorimer} \\
B1800-27    & 0.334415426505           & 406.781            & 0.143647  & He & \cite{1992Johnston} \\
J1810-2005  & 0.032822244860022        & 15.01201911        & 0.279042  & He(?) & \cite{2010Janssen} \\
J1822-0848  & 2.50451781786            & 286.8303           & 0.324447  & He & \cite{2006Lorimer}\\
J1840-0643  & 0.0355778755             & 937.1              & 0.159268  & He & \cite{2013Knispel}\\
J1904+0412  & 0.0710948973807          & 14.934263          & 0.219735  & He & \cite{2001Camilo}\\
J2016+1948  & 0.064940388241514        & 635.02377864       & 0.290699  & He & \cite{2011Gonzalez}\\
B1310+18    & 0.033163166              & 255.8              & 0.298595  & ? & \cite{1991Kulkami}\\
B1718-19    & 1.00403745670            & 0.25827386         & 0.115068  & ? & \cite{2004Hobbs}\\
\hline  
\end{tabular}
\end{table*}

In the ATNF Pulsar Catalogue there are 11 binary pulsar systems with $P_{\rm spin} > 30\,\rm ms$ in which the companions are consistent with He WDs. These systems are listed in Table~\ref{pulsartable3} and plotted in Fig.~\ref{nonmsphewds}. Nine of the systems in Table~\ref{pulsartable3} have spins just outside the MSP realm and two have spins of over a second, typical of normal pulsars. The 9 mildly recycled pulsar systems are plotted in purple in Fig.~\ref{nonmsphewds} and the 2 normally-spinning pulsar systems are plotted in green. The triangles indicate two systems with unknown companion type.  Fig.~\ref{nonmsphewds} shows that there are mildly recycled pulsars all along the $M_{\rm c}$--$P_{\rm orb}$ relation but there are more at higher periods. This is consistent with the consequences of a CEW. Note that the masses plotted in Fig.~\ref{nonmsphewds} are minimum masses so these systems in reality have actual masses further to the right on the diagram. Even though the two normally-spinning pulsars have not been spun up via RLOF, their companion stars could still fit in this scenario if they lost their envelopes by a CEW.

\section{Conclusions}
Stellar models produce a well-defined $M_{\rm c}$--$P_{\rm orb}$ relation for binary millisecond pulsars with He WD companions. It extends to periods of less than $1\,\rm d$ for a $1\,\rm M_{\odot}$ donor. We have explored the $M_{\rm c}$--$P_{\rm orb}$ relation further by studying the range of initial orbital periods required to produce the MSPs with He WD companions observed in our Galaxy. We found that the canonical $M_{\rm c}$--$P_{\rm orb}$ relation \citep{2014Smedley} gives a wide range of post-RLOF orbital periods for a very narrow range in initial orbital periods at the onset of RLOF. Such fine tuning would reduce the likelihood that such systems are produced. Taking inspiration from extraordinary systems such as Z~Her, a companion enhanced wind \citep{1987Tout} was applied to the formation scenario of MSPs with He WD companions. The aim was to discover whether an increased mass-loss rate owing to the presence of a companion can reduce the fine tuning at short periods. We list our main conclusions.
\par Our models indicate that the fine tuning at short orbital periods can be significantly reduced if a CEW is active. We calculated the final period distributions for $B=0$ and $1000$ for the case in which the NS retains all of the mass that is transferred. The two cases give similar final period distributions but there are $23$ times more systems in the final period range of $1$ to $10\,\rm d$ for $B=1000$ than for $B=0$. So when a CEW is operational, more systems are made at shorter periods and this relaxes the fine tuning of the $M_{\rm c}$--$P_{\rm orb}$ relation.
\par Increasing the strength of the CEW decreases the range of initial periods that produce MSPs. MSPs are only produced if RLOF occurs and enough matter is accreted by the NS to spin it up to millisecond periods. As $B$ increases, more mass is lost from the donor in a CEW and, if the companion loses its envelope before the two stars are close enough for RLOF to occur, the pulsar does not spin up. Systems with a high $B$ cannot produce the longer final periods that are seen in nature. A wide range of $B$ is needed to produce the observed population. This reinforces the delicate interplay between the processes involved in MSP production. 
\par Decreasing $\beta$, the fraction of transferred mass retained by the pulsar, keeps the $P_{\rm orb,i}$--$P_{\rm orb,f}$ relation finely tuned at low core masses even with increasing $B$. 
\par The $M_{\rm c}$--$P_{\rm orb}$ relation for MSPs with He WD companions is pushed to lower masses if a CEW is left operational after the cessation of RLOF owing to additional stripping of the collapsing white dwarf envelopes. The CEW intrinsically predicts that mildly recycled pulsars should be present along the entire $M_{\rm c}$--$P_{\rm orb}$ relation for MSPs with He WD companions. Eleven systems listed in the ATNF catalogue are consistent with this prediction but there are not yet enough systems to make a statistically significant statement. 
\par The two systems in Table~\ref{pulsartable3} with $P_{\rm spin} > 1\,\rm s$ could be systems with initial periods above $P_{\rm i,crit}$ for their $B$ or could have formed via a different formation channel.

\section*{Acknowledgements}
SLS thanks STFC for her studentship and her STEP award. CAT thanks Churchill College for his fellowship.


\label{lastpage}

\begin{thebibliography}{99}


\bibitem[{{Alexander} \& {Ferguson}(1994)}]{1994Alex}
{Alexander} D.~R., {Ferguson} J.~W., 1994, ApJ, 437, 879

\bibitem[{{Alpar} {et~al}\mbox{.}(1982){Alpar}, {Cheng}, {Ruderman}, \&
  {Shaham}}]{1982Alpar}
{Alpar} M.~A., {Cheng} A.~F., {Ruderman} M.~A., {Shaham} J., 1982, Nat, 300,
  728

\bibitem[{{Angulo} {et~al}\mbox{.}(1999){Angulo}, {Arnould}, {Rayet},
  {Descouvemont}, {Baye}, {Leclercq-Willain}, {Coc}, {Barhoumi}, {Aguer},
  {Rolfs}, {Kunz}, {Hammer}, {Mayer}, {Paradellis}, {Kossionides}, {Chronidou},
  {Spyrou}, {degl'Innocenti}, {Fiorentini}, {Ricci}, {Zavatarelli},
  {Providencia}, {Wolters}, {Soares}, {Grama}, {Rahighi}, {Shotter}, \& {Lamehi
  Rachti}}]{1999Angulo}
{Angulo} C. {et~al.}, 1999, Nuclear Physics A, 656, 3

\bibitem[\protect\citeauthoryear{Antoniadis et al.}{2013}]{antoniadis2013}
Antoniadis, J., Freire, P. C. C., Wex, N., Tauris, T. M., Lynch, R. S., van Kerkwijk, M. H., Kramer, M., Bassa, C., Dhillon, V. S., Driebe, T., Hessels, J. W. T., Kaspi, V. M., Kondratiev, V. I., Langer, N., Marsh, T. R., McLaughlin, M. A., Pennucci, T. T., Ransom, S. M., Stairs, I. H., van Leeuwen, J., Verbiest, J. P. W. and Whelan, D. G., 2013, Science, 340, 448


\bibitem[{{Bhattacharya \& van der Heuvel}\mbox{.}(1991)}] {1991Bhatt} {{Bhattacharya}, D. and {van den Heuvel}, E.~P.~J.}, 1991, Phys. Rep., 203, 1-124

\bibitem[{{B{\"o}hm-Vitense}(1958)}]{1958Bohm}
{B{\"o}hm-Vitense} E., 1958, Z. Astrophys., 46, 108

\bibitem[{{Buchler} \& {Yueh}(1976)}]{1976Buchler}
{Buchler} J.~R., {Yueh} W.~R., 1976, ApJ, 210, 440

\bibitem[{{Burderi} {et~al}\mbox{.}(2005){Burderi}, {D'Antona}, {di
      Salvo}, {Lavagetto}, {Iaria}, \& {Robba}}]{spinup} {Burderi} L.,
  {D'Antona} F., {di Salvo} T., {Lavagetto} G., {Iaria} R., {Robba}
  N.~R., 2005, in {Rasio} F.~A., {Stairs} I.~H., eds, ASP
  Conf. Ser. Vol.~328, Binary Radio Pulsars. Astron. Soc. Pac., San
  Francisco, p.~269

\bibitem[\protect\citeauthoryear{Camilo et~al.}{2001}]{2001Camilo}
Camilo, F., Lyne, A. G., Manchester, R. N., Bell, J. F., Stairs, I. H., D'Amico, N., Kaspi, V. M., Possenti, I., Crawford, F. and McKay, N. P. F., 2001, ApJ, 548, L187

\bibitem[{{Canuto}(1970)}]{1970Can}
{Canuto} V., 1970, ApJ, 159, 641

\bibitem[{{Caughlan} \& {Fowler}(1988)}]{1988Cau}
{Caughlan} G.~R., {Fowler} W.~A., 1988, Atomic Data and Nuclear Data Tables,
  40, 283

\bibitem[\protect\citeauthoryear{Edwards \& Bailes}{2001}]{2001Edwards}Edwards, R. T. and Bailes, M., 2001, ApJ, 553, 801

\bibitem[{{Eggleton}(1971)}]{1971Eggleton}
{Eggleton} P.~P., 1971, MNRAS, 151, 351

\bibitem[{{Ferguson} {et~al}\mbox{.}(2005){Ferguson}, {Alexander}, {Allard},
  {Barman}, {Bodnarik}, {Hauschildt}, {Heffner-Wong}, \& {Tamanai}}]{2005Ferg}
{Ferguson} J.~W., {Alexander} D.~R., {Allard} F., {Barman} T., {Bodnarik}
  J.~G., {Hauschildt} P.~H., {Heffner-Wong} A., {Tamanai} A., 2005, ApJ, 623,
  585

\bibitem[{{Ferrario et~al}\mbox{.}(2007)}] {2007Lilia} {{Ferrario}, L. and {Wickramasinghe}, D.}, 2007, MNRAS, 375, 1009

\bibitem[{{Gonzalez} {et~al}\mbox{.}(2011){Gonzalez}, {Stairs}, {Ferdman},
  {Freire}, {Nice}, {Demorest}, {Ransom}, {Kramer}, {Camilo}, {Hobbs},
  {Manchester}, \& {Lyne}}]{2011Gonzalez}
{Gonzalez} M.~E. {et~al.}, 2011, ApJ, 743, 102


\bibitem[\protect\citeauthoryear{Hobbs et~al.}{2004}]{2004Hobbs} Hobbs, G., Lyne, A. G., Kramer, M., Martin, C. E. and Jordan, C., 2004, MNRAS, 353, 1311

\bibitem[{{Hubbard} \& {Lampe}(1969)}]{1969Hub}
{Hubbard} W.~B., {Lampe} M., 1969, ApJS, 18, 297

\bibitem[{{Hurley} {et~al}\mbox{.}(2010) {Hurley}, {Tout}, {Wickramasinghe}, {Ferrario}, \& {Kiel}}]{2010Hurley} {Hurley}, J.~R.,
 {Tout}, C.~A., {Wickramasinghe}, D.~T., {Ferrario}, L. and {Kiel}, P.~D., 2010, MNRAS, 402, 1437


\bibitem[{{Iben} \& {Renzini}(1983)}]{1983Iben}
{Iben}, I.~Jr., {Renzini}, A., 1983, ARA\&A, 21, 271 

\bibitem[{{Iglesias} \& {Rogers}(1996)}]{1996Iglesias}
{Iglesias} C.~A., {Rogers} F.~J., 1996, ApJ, 464, 943

\bibitem[{{Illarionov} \& {Sunyaev}(1975)}]{1975Illarionov}{{Illarionov}, A.~F. and {Sunyaev}, R.~A.}, 1975, A\&A, 39, 185

\bibitem[{{Ivanova} {et~al}\mbox{.}(2013)}]{2013Ivanova} {Ivanova}, N., {Justham}, S., {Chen}, X., {De Marco}, O., {Fryer}, C.~L., {Gaburov}, E., {Ge}, H., {Glebbeek}, E., {Han}, Z., {Li}, X.-D., {Lu}, G., {Marsh}, T., {Podsiadlowski}, P., {Potter}, A., {Soker}, N., {Taam}, R., {Tauris}, T.~M., {van den Heuvel}, E.~P.~J. and {Webbink}, R.~F., 2013, A\&A, 21, 59


\bibitem[\protect\citeauthoryear{Jacoby et~al.}{2005}]{2005Jacoby}
{Jacoby} B.~A., {Hotan} A.~W., {Bailes} M., {Ord} S.~M., {Kulkarni} S.~R.,
  2005, BAAS, 37, 1468

\bibitem[\protect\citeauthoryear{Janssen et~al.}{2010}]{2010Janssen} Janssen, G. H., Stappers, B. W., Bassa, C. G., Cognard, I., Kramer, M. and Theureau, G., 2010, A\&A, 514, A74

\bibitem[{{Jia} \& {Li}(2014)}]{2014Jia}
{Jia} K.~A. and {Li} X.~-D., 2014, ApJ, 791, 127

\bibitem[\protect\citeauthoryear{Johnston et~al.}{1992}]{1992Johnston}Johnston, S., Manchester, R. N., Lyne, A. G., Bailes, M., Kaspi, V. M., Qiao, G. and D'Amico, N., 1992, ApJ, 387, L37


\bibitem[\protect\citeauthoryear{Knispel et~al.}{2013}]{2013Knispel} Knispel, B., Eatough, R. P., Kim, H., Keane, E. F., Allen, B., Anderson, D., Aulbert, C., Bock, O., Crawford, F., Eggenstein, H.-B., Fehrmann, H., Hammer, D., Kramer, M., Lyne, A. G., Machenschalk, B., Miller, R. B., Papa, M. A., Rastawicki, D., Sarkissian, J., Siemens, X. and Stappers, B. W., 2013, ApJ, 774, 93

\bibitem[\protect\citeauthoryear{Kulkami et~al.}{1991}]{1991Kulkami} Kulkarni, S. R., Anderson, S. B., Prince, T. A. and Wolszczan, A., 1991, Nature, 349, 47

\bibitem[{{Landau} \& {Lifshitz}(1959)}]{1959Landau} {{Landau}, L.~D. and {Lifshitz}, E.~M.}, 1959, {Course of theoretical physics, Oxford: Pergamon Press}


\bibitem[\protect\citeauthoryear{Lorimer et~al.}{2006}]{2006Lorimer} Lorimer, D. R., Faulkner, A. J., Lyne, A. G., Manchester, R. N., Kramer, M., McLaughlin, M. A., Hobbs, G., Possenti, A., Stairs, I. H., Camilo, F., Burgay, M., D'Amico, N., Corongiu, A. and Crawford, F., 2006, MNRAS, 372, 777

\bibitem[{{Paczy{\'n}ski}(1971)}]{pacz}
{Paczy{\'n}ski} B., 1971, Acta Astron., 21, 417

\bibitem[\protect\citeauthoryear{Pols et~al.}{1995}]{1995Pols}
{Pols} O.~R., {Tout} C.~A., {Eggleton} P.~P., {Han} Z., 1995,
  MNRAS, 274, 964

\bibitem[{{Radhakrishnan} \& {Srinivasan}(1982)}]{1982Radhak}
{Radhakrishnan} V., {Srinivasan} G., 1982, Current Science, 51, 1096

\bibitem[{{Refsdal} \& {Weigert}(1971)}]{1971Refsdal}
{Refsdal} S., {Weigert} A., 1971, AAP, 13, 367

\bibitem[{{Reimers}(1975)}]{1975Reimers}
{Reimers} D., 1975, Mem. Soc. R. Sci, 8, 369

\bibitem[\protect\citeauthoryear{Schr\"oder, Pols \& Eggleton}{1997}]{1997Sch}
{Schr\"oder} K.-P., {Pols} O.~R., {Eggleton} P.~P., 1997, MNRAS, 285, 696

\bibitem[\protect\citeauthoryear{Schwab, Podsiadlowski \& Rappaport}{2010}]{2010Schwab}
{Schwab} J., {Podsiadlowski} P. and {Rappaport} S., 2010, ApJ, 719, 722

\bibitem[{{Smedley et~al}\mbox{.}(2014)}]{2014Smedley}{{Smedley}, S.~L. and {Tout}, C.~A. and {Ferrario}, L. and {Wickramasinghe}, D.~T.}, 2014, MNRAS, 437, 2217

\bibitem[\protect\citeauthoryear{Smith}{1979}]{smith1979}Smith M. A.,
  1979, PASP, 91, 737

\bibitem[{{Splaver}(2004)}]{2004Splaver}
{Splaver} E.~M., 2004, PhD thesis, Princeton Univ.

\bibitem[{{Splaver} {et~al}\mbox{.}(2005){Splaver}, {Nice}, {Stairs}, {Lommen},
  \& {Backer}}]{2005Splaver}
{Splaver} E.~M., {Nice} D.~J., {Stairs} I.~H., {Lommen} A.~N., {Backer} D.~C.,
  2005, ApJ, 620, 405

\bibitem[{{Stancliffe} \& {Eldridge}(2009)}]{2009Stan}
{Stancliffe} R.~J., {Eldridge} J.~J., 2009, MNRAS, 396, 1699

\bibitem[{{Tout} \& {Eggleton}(1988)}]{1987Tout}
{Tout} C.~A. and {Eggleton} P.~P., 1988, MNRAS, 231, 823

\bibitem[{{Tout} \& {Hall}(1991)}]{1991Tout}
{Tout} C.~A. and {Hall} D.~S., 1991, MNRAS, 253, 9

\bibitem[{{Verbunt} \& {Zwaan}(1981)}]{1981Verbunt}{Verbunt} F. and {Zwaan} C., 1981, A\&A, 100, L7

\bibitem[{{Verbiest} {et~al}\mbox{.}(2008){Verbiest}, {Bailes}, {van Straten},
  {Hobbs}, {Edwards}, {Manchester}, {Bhat}, {Sarkissian}, {Jacoby}, \&
  {Kulkarni}}]{2008Verbiest}
{Verbiest} J.~P.~W. {et~al.}, 2008, ApJ, 679, 675



\end{thebibliography}
\end{document}